\def\beq{\begin{equation}}
\def\eeq{\end{equation}}
\def\balll{\begin{array}{lll}}
\def\ea{\end{array}}
\def\beqa{\begin{eqnarray}}
\def\eeqa{\end{eqnarray}}

\documentclass[floatfix,twocolumn,aps,prd,showpacs,amsmath,amssymb]{revtex4}
\usepackage{epsfig}
\begin{document}
\title{Low-frequency absorption cross section of the electromagnetic waves for extreme
Reissner-Nordstr\"om black holes in higher dimensions}

\author{Lu\'\i s\ C.\ B.\ Crispino}
\email{crispino@ufpa.br}
\affiliation{Faculdade de F\'\i sica, Universidade Federal do
Par\'a, 66075-110, Bel\'em, PA,  Brazil}

\author{Atsushi Higuchi}
\email{ah28@york.ac.uk}
\affiliation{Department of Mathematics, University of York, Heslington,
York YO10 5DD, United Kingdom}

\author{George\ E.\ A.\ Matsas}
\email{matsas@ift.unesp.br}
\affiliation{Instituto de F\'\i sica Te\'orica, Universidade Estadual Paulista,
Rua Dr.\ Bento Teobaldo Ferraz, 271 - Bl.\ II, 01140-070, S\~ao Paulo, SP,  Brazil}

\date{August 31, 2010}
\begin{abstract}
We investigate the low-frequency absorption cross section of the electromagnetic waves for extreme Reissner-Nordstr\"om black holes in
higher dimensions.  We first construct the exact solutions to the relevant wave equations in the zero-frequency limit.
In most cases it is possible to use these solutions to find the transmission coefficients of partial waves in the low-frequency limit.  We use these transmission coefficients to calculate the low-frequency absorption cross section in five and six spacetime dimensions.
We find that this cross section is dominated by the modes with $\ell=2$ in the spherical-harmonic expansion
rather than those with $\ell=1$, as might have been expected, because of
the mixing between the electromagnetic and gravitational waves.  We also find an upper limit for the low-frequency absorption cross section in
dimensions higher than six.
\end{abstract}

\maketitle

\section{Introduction}

Motivated by proposals of gravity theories in the context of extra dimensions,
the absorption cross section of spherically symmetric black holes
in higher dimensions has been investigated for different fields.
The scalar case has been analyzed in Refs.~\cite{gibbons, scalarfermion},
the fermionic case in Refs.~\cite{gibbons, scalarfermion, fermion},
and the gravitational case in Ref.~\cite{grav}.
The massless scalar case has also been investigated by one of the present authors
for stationary black holes~\cite{atsushi}, whereas the massive scalar case has been
studied for Reissner-Nordstr\"om black holes in Ref.~\cite{mscalar}.
Investigations for the scalar field in charged black hole spacetimes can
also be found in Ref.~\cite{jungetalnpb}.

Sometime ago the present authors studied the behavior of the absorption cross section of the massless vector field for
extreme Reissner-Nordstr\"om black holes in arbitrary dimensions~\cite{CHM},
generalizing a result of Gubser in four dimensions~\cite{Gubser}.
However, the massless vector field equation in the Reissner-Nordstr\"om
background does not describe the electromagnetic field because of the mixing between the electromagnetic and
gravitational waves~\cite{Gerlach,Zerilli,Moncrief,Olson-Unruh,Matzner}.  In Ref.~\cite{CHO},
which followed up the work in Ref.~\cite{CO}, the
absorption cross section of electromagnetic waves for Reissner-Nordstr\"om black holes in four dimensions was studied, and for the extreme case
it was found that some modes with $\ell=2$ in the spherical harmonic expansion contributed at the same order of the frequency $\omega$ as the modes with $\ell=1$ because of the mixing between the electromagnetic and gravitational waves.

In this paper we study the low-frequency behavior of the absorption cross section of the electromagnetic waves for the higher-dimensional extreme
Reissner-Nordstr\"om black holes.  In particular, we determine the low-frequency behavior of the cross section in five and six spacetime
dimensions and find
that it is dominated by some modes with $\ell=2$ rather than those with $\ell=1$.  The rest of this paper is organized as follows.  In Sec.~\ref{exact}
we present the exact solutions to the equations describing the system of electromagnetic and gravitational waves found by
 Kodama and Ishibashi~\cite{Ishibashi-Kodama} in the zero-frequency limit.  In Sec.~\ref{low-freq} we discuss
the low-frequency behavior of the transmission coefficients of partial waves. In Sec.~\ref{absorptionCS} we find the low-frequency absorption cross
section in dimensions five and six and its upper bound in dimensions higher than six.
In Sec.~\ref{summary} we summarize our results and make some concluding remarks.

\section{Exact solutions to the zero-frequency equations} \label{exact}

As in four dimensions, the modes for the electromagnetic waves and the part of the gravitational waves that mix with the electromagnetic waves
have angular dependence given either by vector or scalar spherical harmonics.  Following Chandrasekhar~\cite{Chandrasekhar}, we call the modes described by vector spherical harmonics the axial modes and those described by the scalar spherical harmonics the polar modes. (In Ref.~\cite{Ishibashi-Kodama} they are called the vector-type and scalar-type perturbations, respectively. They are also called the even- and
odd-parity
perturbations, respectively.) The line element of the $p+2$ dimensional extreme
Reissner-Nordstr\"om black hole (with $p \geq 2$) is given by~\cite{Myers-Perry}
\beq
ds^2 = -f(r)dt^2
+ \left[f(r)\right]^{-1}dr^2 + r^2d\Omega^2, \label{metric}
\eeq
where $d\Omega^2$ is the line element of the unit $p$-dimensional sphere, and where
\beq
f(r) = \left[ 1- \left(\frac{r_H}{r}\right)^{p-1}\right]^2.
\eeq

Let us first present the radial wave equations for the modes describing the electromagnetic and gravitational perturbations in this spacetime, which are special cases of the general equations found in Ref.~\cite{Ishibashi-Kodama}.  The radial functions $\Phi_{a\pm}(r)$ describing the axial perturbations satisfy
\beq
\left[f(r)\frac{d\ }{dr}f(r)\frac{d\ }{dr} + \omega^2 - \frac{f(r)}{r^2}V_{a\pm}(r)\right]\Phi_{a\pm}(r) = 0, \label{axialEq}
\eeq
where
\beqa
V_{a\pm} & = & \left(\ell+\frac{p}{2}-1\right)\left(\ell+\frac{p}{2}\right) + \frac{p(5p-2)}{4}\left(\frac{r_H}{r}\right)^{2p-2}\nonumber \\
&& + \left(- \frac{p^2+2}{2} \pm \Delta \right)\left(\frac{r_H}{r}\right)^{p-1},\\
\Delta & \equiv & \sqrt{(p^2-1)^2 + 2p(p-1)(\ell-1)(\ell+p)}.
\eeqa
We have $\ell \geq 1$ for $\Phi_{a+}$ whereas $\ell \geq 2$ for $\Phi_{a-}$.
These functions are given in terms of the radial functions $\Phi_{ae}$ and $\Phi_{ag}$ which describe the electromagnetic and gravitational
axial perturbations, respectively, by
\beqa
\Phi_{a+} & = & \cos\psi^{(a)}_{\ell} \Phi_{ae} + \sin\psi^{(a)}_{\ell} \Phi_{ag},\\
\Phi_{a-} & = & -\sin\psi^{(a)}_{\ell} \Phi_{ae} + \cos\psi^{(a)}_{\ell}\Phi_{ag},
\eeqa
where (with $|\psi_\ell^{(a)}| < \pi/4$)
\beq
\sin 2\psi^{(a)}_\ell = \sqrt{\frac{2p(\ell-1)(\ell+p)}{2p(\ell-1)(\ell+p)+(p-1)(p+1)^2}}.  \label{axialmix}
\eeq
The radial functions $\Phi_{p\pm}(r)$ describing the polar perturbations satisfy
\beq
\left[f(r)\frac{d\ }{dr}f(r)\frac{d\ }{dr} + \omega^2 - \frac{f(r)}{r^2}V_{p\pm}\right]\Phi_{p\pm}(r) = 0,  \label{polarEq}
\eeq
where, with the definition
\beq
\xi \equiv (r_H/r)^{p-1}, \label{xi}
\eeq
we have
\beqa
V_{p+} & = & \frac{1}{4(p\xi-\ell-p)^2}\left\{ (\ell+p)^2(2\ell+p-2)(2\ell+p)\right.\nonumber \\
&& -2(\ell+p)(4\ell^2+3p^2\ell-4p\ell-2p^2)\xi\nonumber \\
&& +\left[(p^2-6p+8)\ell^2 - 6p^3\ell - 6p^4\right]\xi^2\nonumber \\
&& + \left[(-4p^2+6p^3)\ell + 8p^4-4p^3\right]\xi^3\nonumber \\
&& \left. + (2p^3-3p^4)\xi^4\right\}.
\eeqa
Interestingly, the potential $V_{p-}$ is obtained from $V_{p+}$ by letting $\ell \to -\ell-p+1$.
We have $\ell\geq 1$ for $\Phi_{p+}$ and $\ell \geq 2$ for $\Phi_{p-}$.
These functions are given in terms of the radial functions $\Phi_{pe}$ and $\Phi_{pg}$ which describe the electromagnetic and gravitational
polar perturbations, respectively, by
\beqa
\Phi_{p+} & = & \cos\psi^{(p)}_{\ell} \Phi_{pe} + \sin\psi^{(p)}_{\ell} \Phi_{pg},\\
\Phi_{p-} & = & -\sin\psi^{(p)}_{\ell} \Phi_{pe} + \cos\psi^{(p)}_{\ell}\Phi_{pg},
\eeqa
where (with $|\psi_\ell^{(p)}| < \pi/4$)
\beq
\sin 2\psi^{(p)}_\ell = \sqrt{\frac{4(\ell-1)(\ell+p)}{4(\ell-1)(\ell+p)+(p+1)^2}}. \label{polarmix}
\eeq
A more useful form is
\beq
\sin^2\psi^{(p)}_\ell = \frac{\ell-1}{2\ell+p-1}. \label{polarmix2}
\eeq

In all cases the zero-frequency radial equation with $\omega=0$ can be written in terms of the variable $\xi$ defined by Eq.~(\ref{xi}) as
\beqa
&& \left[ \xi^2(1-\xi)^2\frac{d^2\ }{d\xi^2} + \xi(1-\xi)\left( 1-3\xi + \frac{1-\xi}{p-1}\right)\frac{d\ }{d\xi}\right.\nonumber \\
&& \,\,\,\,\,\,\,\,\,\,\,\,\,\,\,\,\,\,\,\,\, \left. - \frac{V(\xi)}{(p-1)^2}\right]\Phi = 0, \label{generalEq}
\eeqa
where $V(\xi)$ is $V_{a\pm}$ or $V_{p\pm}$, and $\Phi$ is $\Phi_{a\pm}$ or $\Phi_{p\pm}$.

For the axial case Eq.~(\ref{generalEq}) can be transformed to the standard hypergeometric equation by multiplying $\Phi$ by an appropriate factor.
We find in this manner the following solutions:
\beqa
\Phi^{(1)}_{a\pm} & = & \xi^{-(2\ell+p)/\left[2(p-1)\right]}(1-\xi)^{\lambda^{(1)}_\pm}\nonumber \\
  &\times& F(\lambda^{(1)}_{\pm}-\tfrac{\ell+p}{p-1},\lambda^{(1)}_{\pm}-\tfrac{\ell-2p+1}{p-1}; 2(\lambda^{(1)}_{\pm}+1);1-\xi), \nonumber \\
\label{axial-relevant}
\eeqa
where
\beq
\lambda^{(1)}_{\pm} =  - \frac{1}{2} + \sqrt{\frac{1}{4} + \frac{1}{(p-1)^2}\left(p^2-1+(\ell-1)(\ell+p)\pm \Delta\right)}. \label{beforecomment}
\eeq
Since $p^2-1+(\ell-1)(\ell+p)\pm \Delta$ are increasing functions of $\ell$ for $\ell\geq 1$ and are
non-negative for $\ell=1$, the constants $\lambda^{(1)}_{\pm}$ are both real.
The other independent solutions can be chosen as
\beqa
\Phi^{(2)}_{a\pm} & = & \xi^{-(2\ell+p)/\left[2(p-1)\right]}(1-\xi)^{\lambda^{(2)}_\pm}\nonumber \\
& \times& F(\lambda^{(2)}_{\pm}-\tfrac{\ell+p}{p-1},\lambda^{(2)}_{\pm}-\tfrac{\ell-2p+1}{p-1}; 2(\lambda^{(2)}_{\pm}+1);1-\xi),\nonumber \\ \label{second}
\eeqa
where
\beq
\lambda^{(2)}_{\pm} =  - \frac{1}{2} - \sqrt{\frac{1}{4} + \frac{1}{(p-1)^2}\left(p^2-1+(\ell-1)(\ell+p)\pm \Delta\right)},
\eeq
if $2\lambda^{(2)}_{\pm}$ is not an integer. If it is an integer, the solutions $\Phi^{(2)}_{a\pm}$ are not valid, but valid solutions can be
generated by the standard method, and they
behave like $(1-\xi)^{\lambda^{(2)}_\pm}$ for $|1-\xi|\ll 1$.

Now, let us examine the radial equations for the polar perturbations.  Remarkably, the following functions are solutions to Eq.~(\ref{generalEq})
with $V(\xi) = V_{p+}(\xi)$:
\beqa
\Phi^{(1)}_{p+} & = & \xi^{-(2\ell+p)/\left[2(p-1)\right]}(1-\xi)^{(\ell+p-1)/(p-1)}\nonumber \\
&& \times \frac{1}{\ell+p-p\xi}, \label{polarplus1}\\
\Phi^{(2)}_{p+} & = & \xi^{(2\ell+p-2)/\left[2(p-1)\right]}(1-\xi)^{-(\ell+2p-2)/(p-1)}\nonumber \\
&& \times \left(\frac{1}{\ell+p-p\xi} + a_{\ell+} + b_{\ell+}\xi\right),
\eeqa
where
\beqa
a_{\ell+} & = & \frac{3p}{\ell^2} - \frac{(p-2)(p+1)}{(\ell+p)(p-1)^2}+ \frac{2p+3}{\ell(p-1)},\\
b_{\ell+} & = & -\frac{3p}{\ell^2} - \frac{p(p+1)}{(\ell+p)^2(p-1)^2}\nonumber \\
&& + \frac{2p+3}{(\ell+p)(p-1)} - \frac{2p+3}{\ell(p-1)}.
\eeqa
These solutions are linearly independent for all $p \geq 2$ and $\ell \geq 1$.  The solutions with $V(\xi)=V_{p-}(\xi)$ can be obtained from these
by replacing $\ell$ by $-\ell-p+1$ unless $2\ell=p-1$:
\beqa
\Phi^{(1)}_{p-} & = & \xi^{(2\ell+p-2)/\left[2(p-1)\right]}(1-\xi)^{-\ell/(p-1)}\nonumber \\
&& \times \frac{1}{p\xi+\ell-1},\label{polarminus1}\\
\Phi^{(2)}_{p-} & = & \xi^{-(2\ell+p)/\left[2(p-1)\right]}(1-\xi)^{(\ell-p+1)/(p-1)}\nonumber \\
&& \times \left(\frac{1}{p\xi+\ell-1} + a_{\ell-} + b_{\ell-}\xi\right), \label{polarminus2}
\eeqa
where
\beqa
a_{\ell-} & = & - \frac{3p}{(\ell+p-1)^2} - \frac{(p-2)(p+1)}{(\ell-1)(p-1)^2}\nonumber \\
&&  + \frac{2p+3}{(\ell+p-1)(p-1)},\\
b_{\ell-} & = & \frac{3p}{(\ell+p-1)^2} + \frac{p(p+1)}{(\ell-1)^2(p-1)^2}\nonumber \\
&& + \frac{2p+3}{(\ell-1)(p-1)} - \frac{2p+3}{(\ell+p-1)(p-1)}.
\eeqa
If $2\ell=p-1$, $\Phi_{p-}^{(2)}$ is proportional to $\Phi_{p-}^{(1)}$.  In this case an independent second solution is given by
\beqa
\Phi_{p-}^{(2')} & = & \xi^{-(2p-1)/\left[2(p-1)\right]}(1-\xi)^{-\frac{1}{2}}\frac{1}{2p\xi+p-3}\nonumber \\
&&\times \left[-\frac{1}{2}(p-3)^2-(5p-3)(p-3)\xi\right.\nonumber \\
&& \left. + 9(p-1)^2\xi^2\log \frac{\xi}{1-\xi}\right].
\eeqa

\section{Transmission coefficients for low frequencies} \label{low-freq}

The wave equations (\ref{axialEq}) and (\ref{polarEq}) take the form
\beq
\left[ \frac{d^2\ }{dr_*^2} + \omega^2 - \frac{f(r)}{r^2}V(r)\right]\Phi = 0, \label{generalEq2}
\eeq
where the Regge-Wheeler tortoise coordinate $r_*$ is defined by
\beq
\frac{dr_*}{dr} = \left[ 1- \left(\frac{r_H}{r}\right)^{p-1}\right]^{-2}.
\eeq
We note that
\beqa
r_* & \sim & r \,\,\,{\rm if}\,\,\,\,r \gg r_H,\\
r_* & \sim  & -\frac{r_H^2}{(p-1)^2(r-r_H)}\,\,\,{\rm if}\,\,\, r-r_H \ll r_H.
\eeqa
In all cases, we have
\beq
f(r)V(r) \approx \left(\ell+\frac{p}{2}-1\right)\left(\ell+\frac{p}{2}\right),\,\,\,r\gg r_H.
\eeq
Hence the large $r$ behavior of the solutions relevant to the absorption process is given by
\beqa
\Phi & \approx & \Phi_{r\gg r_H}\nonumber \\
& = &  \sqrt{\frac{\pi\omega r}{2}}
\left[ H^{(2)}_{\ell+\frac{p-1}{2}}(\omega r) + (1-b_\ell(\omega))
H^{(1)}_{\ell+\frac{p-1}{2}}(\omega r)\right],\nonumber \\
\label{large-rbehavior}
\eeqa
where $b_\ell (\omega)$ is some smooth complex function.
Near the horizon, i.e.\ for $r_*$ large and negative, Eq.~(\ref{generalEq2}) can be approximated by
\beq
\left[\frac{d^2\ }{dr_*^2} + \omega^2 - \frac{\nu_\ell^2-\frac{1}{4}}{r_*^2}\right]\Phi \approx 0,
\label{equationnearhorizon}
\eeq
where
\beq
\nu_\ell = \sqrt{\frac{V(r_H)}{(p-1)^2}+\frac{1}{4}}.  \label{nuell}
\eeq
Hence, near the horizon we have
\beq
\Phi \approx \Phi_{r \approx r_H} =  D_\ell(\omega) \sqrt{\frac{-\pi\omega r_*}{2}}
H_{\nu_\ell}^{(1)}(-\omega r_*).  \label{near-horizon}
\eeq
Note that $\Phi_{r\gg r_H} \approx  e^{-i\omega r +i\gamma_\ell} + (1-b_\ell(\omega))e^{i\omega r- i\gamma_\ell}$,
where $\gamma_\ell= \frac{\pi}{2}(\ell+\frac{p-2}{2})$ and
$\Phi_{r \approx r_H} \approx D_\ell(\omega) e^{-i\omega r_*}$ up to a phase factor for $|\omega r_*| \gg 1$.
Thus, the transmission coefficient is given by
\beq
\mathcal{P}_\ell \equiv |D_\ell(\omega)|^2.
\eeq
We also have by energy conservation
\beq
|D_\ell(\omega)|^2 = 2{\rm Re}\left[ b_\ell(\omega)\right] - |b_\ell(\omega)|^2. \label{probability}
\eeq

The low-frequency behavior of the transmission coefficients can be found as follows.
Noting that $1-\xi \approx (p-1)(r-r_H)/r_H$ for $r\approx r_H$, we find from Eq.~(\ref{near-horizon})
that for small $\omega$ (with $|\omega r_*| \ll 1$) the solutions behave near the horizon (with $\xi$ fixed) as follows:
\beqa
\Phi_{r\approx r_H} & \approx & -iD_\ell(\omega)\frac{\Gamma(\nu_\ell)}{\sqrt{\pi}}\left[\frac{2(p-1)}{\omega r_H}
\right]^{\nu_\ell-\frac{1}{2}}\left(1-\xi\right)^{\nu_\ell-\frac{1}{2}}\nonumber \\
&& {\rm if}\,\,\,\nu_\ell > 0,
\label{NH-solution1}
\eeqa
and
\beqa
\Phi_{r\approx r_H} & \approx & iD_\ell(\omega)\left[\sqrt{\frac{2\omega r_H}{\pi(p-1)}}\log\omega r_H\right](1-\xi)^{-\frac{1}{2}}\nonumber \\
&& {\rm if}\,\,\, \nu_\ell=0,  \label{NH-solution2}
\eeqa
where we have used the fact that if $-\omega r_* \ll 1$
and $r_* \approx - r_H/[(p-1) (1-\xi)]$, then
$\omega r_H \ll 1-\xi$.
We note that a second independent solution of Eq.~(\ref{equationnearhorizon}) behaves like $(1-\xi)^{-\nu_\ell-\frac{1}{2}}$ if $\nu_\ell>0$ and
$(1-\xi)^{-\frac{1}{2}}\log(1-\xi)$ if $\nu_\ell=0$ near the horizon.  That is, a second
independent solution diverges faster as $\xi\to 1$.  Thus, the $\omega=0$ solution that
diverges more slowly as $\xi\to 1$ corresponds to the $\omega\to 0$ limit of the
solution relevant to the absorption process, and it behaves like $(1-\xi)^{\nu_\ell-\frac{1}{2}}$, where $\nu_\ell$ is given by
Eq.~(\ref{nuell}). Therefore, the relevant solutions are
$\Phi_{a\pm}^{(1)}$ in Eq.~(\ref{axial-relevant}), $\Phi_{p+}^{(1)}$ in Eq.~(\ref{polarplus1}),
$\Phi_{p-}^{(2)}$ in Eq.~(\ref{polarminus2}) if $2\ell> p-1$ and $\Phi_{p-}^{(1)}$ in Eq.~(\ref{polarminus1}) if $2\ell \leq p-1$.

To find the behavior of the solutions for $r \gg r_H$ (i.e.\ for $\xi \ll 1$) in the limit $\omega \to 0$ we first note that Eq.~(\ref{large-rbehavior})
can be written as
\beqa
\Phi_{r\gg r_H}
 & =  & \sqrt{2\pi\omega r}
\left[\left(1 - \frac{b_\ell(\omega)}{2}\right)J_{\ell+\frac{p-1}{2}}(\omega r)\right.\nonumber \\
&& \left. \,\,\,\,\,\,\,\,\,\,\,\,\,\,\,\,\,-\frac{ib_\ell(\omega)}{2}N_{\ell+\frac{p-1}{2}}(\omega r)\right].
\label{large-rbehavior2}
\eeqa
We assume that the limit of $\omega^{-2\ell-p+1}|b_\ell(\omega)|$ as $\omega \to 0$ either exists or is
infinite. For small $\omega$ and for small $\xi$, we have
\beqa
\Phi_{r\gg r_H} & \approx &  \frac{\sqrt{4\pi}}{\Gamma\left(\ell+\frac{p+1}{2}\right)}\left(\frac{\omega r_H}{2}\right)^{\ell+\frac{p}{2}}
\xi^{-(2\ell+p)/\left[2(p-1)\right]}\nonumber \\
&& {\rm if}\,\,\, \lim_{\omega\to 0}\omega^{-2\ell-p+1}|b_\ell(\omega)| <\infty.
\label{case1}
\eeqa
It can be seen from Eqs.~(\ref{axial-relevant}), (\ref{polarplus1}), (\ref{polarminus1}) and (\ref{polarminus2}) that
all partial waves except for the ``$-$'' polar modes with $2\ell \leq p-1$ fall into this category.
In each of these cases there is a solution $\Phi_\ell(\xi)$ to the $\omega=0$ equation
such that
\beq
\Phi_\ell(\xi)  \approx \begin{cases} (1-\xi)^{\nu_\ell-\frac{1}{2}} & {\rm for}\,\,\,1-\xi \ll 1\\
 \alpha_\ell\xi^{-(2\ell+p)/\left[2(p-1)\right]} & {\rm for}\,\,\, \xi \ll 1, \end{cases}
\eeq
where $\alpha_\ell$ is a nonzero constant.
By comparing these equations with Eqs.~(\ref{NH-solution1}), (\ref{NH-solution2}) and (\ref{case1}) we find
the low-frequency transmission coefficients as
\beqa
\mathcal{P}_\ell & = & |D_\ell(\omega)|^2\nonumber \\
& = & \frac{4\pi^2}{|\alpha_\ell|^2(p-1)^{2\nu_\ell-1}\left|\Gamma(\nu_\ell)\Gamma(\ell+\frac{p+1}{2})\right|^2}\nonumber \\
&& \times \left(\frac{\omega r_H}{2}\right)^{2(\nu_\ell+\ell)+p-1}\,\,\,{\rm if}\,\,\, \nu_\ell > 0.
\eeqa
We do not need the case $\nu_\ell=0$ here.

The values of $\nu_\ell$ and $\alpha_\ell$ for the axial modes can be found from Eq.~(\ref{axial-relevant}) as
\beqa
\nu_\ell^{(a\pm)} & = & \sqrt{\frac{1}{4} + \frac{1}{(p-1)^2}\left(p^2-1+(\ell-1)(\ell+p)\pm \Delta\right)},\nonumber \\ \label{nueq}\\
\alpha_\ell^{(a\pm)} & = & \frac{\Gamma\left(2(\lambda_{\pm}^{(1)}+1)\right)\Gamma\left(1+\frac{2\ell}{p-1}\right)}
{\Gamma\left(\lambda_{\pm}^{(1)} + \frac{\ell+3p-2}{p-1}\right)\Gamma\left(\lambda_{\pm}^{(1)}+\frac{\ell-1}{p-1}\right)}.
\eeqa
Eq.~(\ref{nueq}) can also be obtained from Eq.~(\ref{nuell}).
The $\nu_\ell^{(a\pm)}$ are increasing as functions of $\ell$ because $(\ell-1)(\ell+p)\pm \Delta$ are increasing.
For the `$+$' polar modes we find the values of $\nu_{\ell}$ and $\alpha_\ell$ from Eq.~(\ref{polarplus1}) as
\beqa
\nu_{\ell}^{(p+)} & = & \frac{\ell}{p-1}+\frac{3}{2},\\
\alpha_{\ell}^{(p+)} & = & \frac{\ell}{\ell+p}.
\eeqa
For the `$-$' polar modes with $2\ell > p-1$  we find from Eq.~(\ref{polarminus2})
\beqa
\nu_{\ell}^{(p-)} & = & \frac{\ell}{p-1} - \frac{1}{2},\\
\alpha_{\ell}^{(p-)} & = & \frac{(\ell-1)(2\ell-p+1)}{(\ell+p-1)(2\ell+p-1)}.
\eeqa

For the `$-$' polar modes with $2\ell \leq p-1$ the $\omega=0$ solutions relevant to the absorption process is
given by Eq.~(\ref{polarminus1}).  We find that they behave like $\xi^{(2\ell+p-2)/\left[2(p-1)\right]}$ rather than
$\xi^{-(2\ell+p)/\left[2(p-1)\right]}$ for small $\xi$.  This apparent difficulty is resolved by noting that
\beqa
\Phi_{r\gg r_H} & \approx & i\frac{b_\ell(\omega)}{\sqrt{\pi}}\left( \frac{2}{\omega r_H}\right)^{\ell+\frac{p}{2}-1}
\nonumber \\
&& \times \Gamma\left(\ell+\tfrac{p-1}{2}\right) \xi^{(2\ell+p-2)/\left[2(p-1)\right]}\nonumber \\
&& {\rm if}\,\,\,
\lim_{\omega\to 0}\omega^{-2\ell-p+1}|b_\ell(\omega)|= \infty.  \label{case2}
\eeqa
Thus, the partial waves for the ``$-$'' polar modes with $2\ell \leq p-1$ fall into this category.  In each of these cases
one can rescale the solutions in Eq.~(\ref{polarminus1}) so that they satisfy
\beq
\Phi_\ell(\xi)  \approx \begin{cases} (1-\xi)^{\nu_\ell-\frac{1}{2}} & {\rm for}\,\,\,1-\xi \ll 1\\
 \alpha_\ell\xi^{(2\ell+p-2)/\left[2(p-1)\right]} & {\rm for}\,\,\, \xi \ll 1 \end{cases}
\eeq
for some $\alpha_\ell$.
The transmission coefficients $\mathcal{P}_\ell = |D_\ell(\omega)|^2$ in these cases satisfy
\beqa
|D_\ell(\omega)|^2
& = & \frac{\left|\Gamma\left(\ell+\frac{p-1}{2}\right)\right|^2|b_\ell(\omega)|^2}
{|\alpha_\ell|^2\left|\Gamma(\nu_\ell)\right|^2(p-1)^{2\nu_\ell-1}}\nonumber \\
&& \times \left(\frac{\omega r_H}{2}\right)^{-2(\ell-\nu_\ell)-p+1}\,\,\,{\rm if}\,\,\, \nu_\ell > 0, \label{nonzeronu}
\eeqa
and
\beqa
|D_\ell(\omega)|^2 & = & \frac{(p-1)\left|\Gamma\left(\ell+\frac{p-1}{2}\right)\right|^2|b_\ell(\omega)|^2}
{4|\alpha_\ell|^2\log^2\omega r_H}\nonumber \\
&& \times \left(\frac{\omega r_H}{2}\right)^{-2\ell-p+1}\,\,\,{\rm if}\,\,\, \nu_\ell=0. \label{zeronu}
\eeqa
We find from Eq.~(\ref{polarminus1}) that for the ``$-$'' polar modes with $2\ell \leq p-1$
\beqa
\nu_{\ell}^{(p-)} & = & \frac{1}{2}-\frac{\ell}{p-1},\\
\alpha_\ell^{(p-)} & = & \frac{p+\ell-1}{\ell-1}.
\eeqa
It is not possible to determine the low-frequency behavior of the transmission coefficients, $\mathcal{P}_\ell^{(p-)}$, in these cases.  However,
it is possible to find their upper limits using Eqs.~(\ref{probability}), (\ref{nonzeronu}) and (\ref{zeronu}).
Thus, for $2\ell \leq p-1$,
\beq
\mathcal{P}_\ell^{(p-)} \leq 4C_\ell^{-1}\left(\frac{\omega r_H}{2}\right)^{2(\ell-\nu_\ell)+p-1},
\eeq
where $C_\ell$ is the coefficient of $|b_\ell(\omega)|^2(\frac{\omega r_H}{2})^{-2(\ell-\nu_\ell)-p+1}$ in Eqs.~(\ref{nonzeronu}) and (\ref{zeronu}).
Here terms that tend to zero faster than $\omega^{2(\ell-\nu_\ell)+p-1}$ as $\omega \to 0$ have been neglected.

\section{Low-frequency absorption cross section} \label{absorptionCS}

Let $\mathcal{P}^{(a\pm)}_\ell$ and $\mathcal{P}^{(p\pm)}_\ell$ be the transmission coefficients of the partial waves in the four different modes.
Then the total absorption cross section of an electromagnetic wave with frequency $\omega$ can be expressed as
\beqa
\sigma & = & \frac{(2\pi)^p}{p\Omega_p \omega^p}\nonumber \\
& \times&
\left\{ \sum_{\ell=1}^\infty
\left[M_\ell^{(p)}\mathcal{P}_\ell^{(p+)}\cos^2\psi^{(p)}_{\ell} + M_\ell^{(a)}\mathcal{P}_\ell^{(a+)}\cos^2\psi^{(a)}_{\ell}\right] \right.\nonumber \\
&& \left. \!\! \sum_{\ell=2}^\infty
\left[M_\ell^{(p)}\mathcal{P}_\ell^{(p-)}\sin^2\psi^{(p)}_{\ell} + M_\ell^{(a)}\mathcal{P}_\ell^{(a-)}\sin^2\psi^{(a)}_{\ell}\right] \right\}, \nonumber \\
\label{absorption}
\eeqa
where
\beq
\Omega_p = \frac{2\pi^{(p+1)/2}}{\Gamma(\frac{p+1}{2})},
\eeq
is the surface area of the $p$-dimensional unit sphere, and where
\beqa
M_\ell^{(p)} & = & \frac{(2\ell+p-1)(\ell+p-2)!}{(p-1)!\ell !},\\
M_\ell^{(a)} & = & \frac{(2\ell+p-1)(\ell+p-1)!}{(\ell+1)(\ell+p-2)(p-2)!(\ell-1)!}
\eeqa
are the multiplicities of the scalar and vector spherical harmonics, respectively, on the $p$ dimensional sphere (see, e.g.\ Refs.~\cite{CH,CM}).  The mixing angles
$\psi^{(a)}_{\ell}$ and $\psi^{(p)}_{\ell}$ are given by Eqs.~(\ref{axialmix}) and (\ref{polarmix2}), respectively.
For $p=2$, $3$ and $4$, the low-frequency of all partial waves have been given in Sec.~\ref{low-freq}, and hence
we can find the low-frequency behavior of the total absorption cross section in these dimensions.  For $p \geq 5$ we can only give
the upper limit of the low-frequency absorption cross section.

We first treat the $p=2$ case, i.e.\ the four-dimensional case,
which has been studied in Ref.~\cite{CHO}.  Since the transmission coefficients $\mathcal{P}_\ell^{(a\pm)}$ and
$\mathcal{P}_\ell^{(p\pm)}$
behave like $\omega^{2(\ell+\nu_\ell) +1}$ at low frequencies, we only need the modes with the lowest value of $\ell+\nu_\ell$.
As is well known~\cite{Chandrasekhar}, we have $\mathcal{P}_\ell^{(a\pm)} = \mathcal{P}_\ell^{(p\pm)}$ for $p=2$.  We indeed find
\beqa
\nu_{\ell}^{(a+)} & = & \nu_{\ell}^{(p+)} = \ell+\frac{3}{2},\\
\nu_{\ell}^{(a-)} & = & \nu_{\ell}^{(p-)} = \ell-\frac{1}{2},\\
\alpha_{\ell}^{(a+)} & = & \alpha_{\ell}^{(p+)} = \frac{\ell}{\ell+2},\\
\alpha_{\ell}^{(a-)} & = & \alpha_{\ell}^{(p-)} = \frac{(\ell-1)(2\ell-1)}{(\ell+1)(2\ell+1)}.
\eeqa
The modes with the lowest value of $\ell+\nu_\ell$ are the `$+$' modes with $\ell=1$ and the `$-$' modes with $\ell=2$.  The transmission
coefficients for these modes are all equal and take the value $\frac{4}{9}(\omega r_H)^8$~\footnote{The transmission coefficients given
by Eq.~(24) of Ref.~\cite{CHO} are the values for electromagnetic waves after taking into account the mixing angles rather than that for the `$\pm$' modes before taking them into account.}. (Interestingly, the transmission coefficients for the `$-$' modes can be obtained from those for the `$+$' modes by reducing the value of $\ell$ by $1$.)  Then from Eq.~(\ref{absorption}) we find
\beq
\sigma|_{{\rm low}\,\omega} = \frac{16\pi r_H^2}{9}(\omega r_H)^6,\,\,\,p=2.
\label{sigma4}
\eeq

For $3 \leq p\leq 4$ we only need to determine the smallest among $\nu_{\ell}^{(a\pm)}+\ell$ and $\nu_{\ell}^{(p\pm)}+\ell$ as stated above.
Since $\nu_{\ell}^{(a\pm)}$ and
$\nu_{\ell}^{(a\pm)}$ are all increasing functions of $\ell$, all we need to find is the smallest of the following four numbers:
\beqa
\nu_1^{(a+)} -1& = & \sqrt{\frac{1}{4}+\frac{2(p+1)}{p-1}}-1,\\
\nu_2^{(a-)} & = & \left[\frac{1}{4}-\frac{\sqrt{(p-1)(p^3+3p^2+3p-1)}}{(p-1)^2}  \right.\nonumber \\
&& \,\,\,\,\,\,\left. + \frac{p^2+p+1}{(p-1)^2}\right]^{\frac{1}{2}},\\
\nu_1^{(p+)}-1 & = & \frac{1}{p-1}+\frac{1}{2},\\
\nu_2^{(p-)} & = & \frac{2}{p-1} - \frac{1}{2}.
\eeqa
In the appendix we prove the following inequalities:
\beq
\nu_2^{(p-)} < \nu_2^{(a-)} < \nu_1^{(p+)}-1 < \nu_1^{(a+)}-1,\,\,\,p\geq 3. \label{inequality}
\eeq
Hence, the low-frequency absorption cross section is dominated by the `$-$' polar modes with $\ell=2$.
Thus, we find the low-frequency absorption cross section as
\beqa
\sigma|_{{\rm low}\,\,\omega} & = & \frac{2\pi (p-1)^2 A_H}{\left[\Gamma\left(\frac{p+3}{2(p-1)}\right)\right]^2}
\left(\frac{\omega r_H}{2(p-1)}\right)^{\frac{4}{p-1}+2},\nonumber \\
&& \,\,\,3\leq p \leq 4, \label{lowishp}
\eeqa
where $A_H = \Omega_p r_H^p$ is the horizon area.

\begin{figure}[ht]
\centering
\epsfig{file=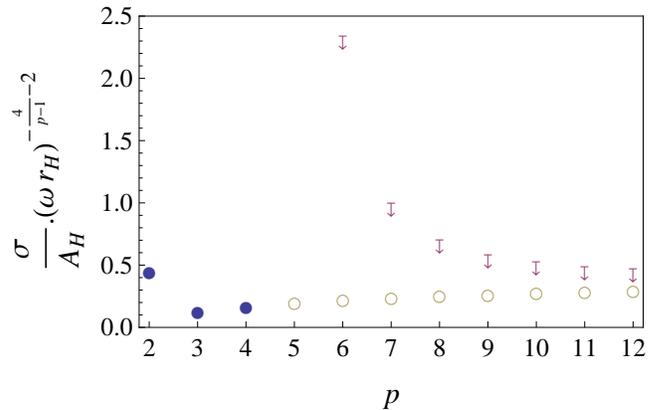,scale=1}
\caption{The low-frequency absorption cross section of electromagnetic waves for extreme
Reissner-Nordstr\"om black holes in $p+2$ dimensions is exhibited. The full circles
represent the exact values obtained for $2 \leq p \leq 4$ as given by Eqs.~(\ref{sigma4})
and~(\ref{lowishp}). The downward arrows indicate the upper limits obtained for $p \geq 6$
as given in Eq.~(\ref{upper2}). The upper limit for $p=5$ is not displayed because of the extra
$\log^2 \omega r_H$ factor, which diverges as  $\omega r_H \to 0$. The conjectured
values for the cross sections with $p \geq 5$, as discussed below Eq.~(\ref{upper2}), are
plotted using empty circles.}
\label{figure}
\end{figure}

For $p\geq 5$ we can only find the upper limit for the low-frequency absorption cross section.  We can use exactly the same argument as above
to conclude that the `$-$' polar modes give the upper bound which tends to zero most slowly as $\omega\to 0$.  We find
\beq
\sigma|_{{\rm low}\,\,\omega} \leq \frac{A_H}{4 \pi}(\omega r_H)^3 \log^2\omega r_H,\,\,\, p=5, \label{upper1}
\eeq
and
\beqa
\sigma|_{{\rm low}\,\,\omega} & \leq &
\frac{2(p-1)^2A_H}{\pi}\left[\Gamma\left(\frac{p-5}{2(p-1)}\right)\right]^2 \nonumber \\
&& \times \left(\frac{\omega r_H}{2(p-1)}\right)^{\frac{4}{p-1}+2},\,\,\, p\geq 6.  \label{upper2}
\eeqa
The upper limit (\ref{upper1}) is clearly consistent with the assumption that the low-frequency absorption cross section is given by
Eq.~(\ref{lowishp}) also for $p=5$.  It is also possible for this equation to be valid for $p\geq 6$ because the right-hand side is
$\cos^2\frac{2\pi}{p-1}$ times the upper limit given by Eq.~(\ref{upper2}).
In Fig.~\ref{figure} we summarize our results for the low-frequency absorption
cross section as a function of $p$.

\section{Summary and discussions} \label{summary}

In this paper we investigated the low-frequency absorption cross section of the electromagnetic waves
for extreme Reissner-Nordstr\"om black holes in higher dimensions using the work of Kodama and Ishibashi~\cite{Ishibashi-Kodama}.
It was found that the low-frequency behavior of the absorption cross section is dominated by modes with $\ell=2$ rather than those with $\ell=1$
due to the mixing between the electromagnetic and gravitational waves in five and six spacetime dimensions, i.e.\ for $p=3$ and $4$. For
dimensions higher than six we only found
upper limits for the low-frequency behavior of the absorption cross sections, which are again dominated by modes with $\ell=2$. These upper limits
are consistent with the assumption that the expression for the low-frequency absorption cross section for dimensions five and six are valid in
dimensions higher than six as well.

In addition our investigation revealed some remarkable features of the equations governing the gravitational and electromagnetic perturbations in the
extreme Reissner-Nordstr\"om background.  Considering the complexity of these equations, particularly of those governing the polar perturbations, it is
already remarkable that they allow simple solutions.  We also noted that the potentials for the two decoupled equations for the polar perturbations are
related by the transformation $\ell\leftrightarrow -\ell-p+1$.  It is likely that these features are only part of a more profound structure in perturbations of higher-dimensional Reissner-Nordstr\"om black holes.  It will be interesting to find whether this is indeed the case.

\acknowledgments

The authors would like to thank Conselho Nacional de Desenvolvimento
Cient\'\i fico e Tecnol\'ogico (CNPq). L.~C.\ and A.~H. also thank
Funda\c{c}\~ao de Amparo \`a Pesquisa do Estado do Par\'a (FAPESPA)
for partial financial support. L.~C.\ would like to acknowledge also
partial financial support from Coordena\c{c}\~ao de Aperfei\c{c}oamento
de Pessoal de N\'\i vel Superior (CAPES). G.~M.\ also thanks Funda\c{c}\~ao de
Amparo \`a Pesquisa do Estado de S\~ao Paulo (FAPESP)
for partial financial support, and
A.~H.\ thanks the International Centre for Theoretical Physics (ICTP)
for a travel grant and
the Universidade Federal do Par\'a (UFPA) in Bel\'em for kind hospitality while
this work was carried out.

\appendix

\section{Proof of inequality (\ref{inequality})}

First we note that
\beqa
&& (p-1)^2\left[(\nu_2^{(a-)})^2 - (\nu_2^{(p-)})^2\right] \nonumber \\
&& = p^2+3p-5 \nonumber \\
&& \,\,\, - \sqrt{(p-1)(p^3+3p^2+3p-1)}.
\eeqa
Since $\nu_2^{(a-)} > 0$ for all $p\geq 2$, the inequality
\beq
p^2 + 3p-5 > \sqrt{(p-1)(p^3+3p^2+3p-1)}
\eeq
will imply that $\nu_2^{(p-)} < \nu_2^{(a-)}$. By squaring both sides and subtracting one from the other,
we find that this inequality is equivalent to
\beq
(p-2)(4p^2+7p-12) > 0,
\eeq
which holds for $p\geq 3$.

Next we note
\beqa
&& (p-1)^2\left[-(\nu_2^{(a-)})^2 + (\nu_1^{(p+)}-1)^2\right]\nonumber \\
&& = \sqrt{(p-1)(p^3+3p^2+3p-1)}-(p^2+1).
\eeqa
This is positive since
\beq
(p-1)(p^3+3p^2+3p-1)-(p^2+1)^2 = 2p(p-2)(p+1)> 0.
\eeq
Since $\nu_1^{(p+)}-1>0$ for $p\geq 2$, this implies that $\nu_2^{(a-)} < \nu_1^{(p+)}-1$.

Finally we find
\beq
(\nu_1^{(a+)})^2 - (\nu_1^{(p+)})^2 = \frac{p-2}{(p-1)^2} > 0,\,\,\,p\geq 3.
\eeq
Since $\nu_1^{(a+)}>0$, this inequality implies $\nu_1^{(p+)} < \nu_1^{(a+)}$.

%###################################################################


\begin{thebibliography}{99}

\bibitem{gibbons} S.\ R.\ Das, G.\ Gibbons,
and S.\ D.\ Mathur, Phys.\ Rev.\ Lett.\ {\bf 78}, 417 (1997).

\bibitem{scalarfermion} E.\ Jung, S.\ H.\ Kim,
and D.\ K.\ Park, J.\ High Energy Phys. {\bf 09} (2004) 005.

\bibitem{fermion} M.\ Rogatko and A.\ Szyplowska,
Phys.\ Rev.\ D {\bf 79}, 104005 (2009).

\bibitem{grav} S.\ Creek, O.\ Efthimiou, P.\ Kanti,
and K.\ Tamvakis, Phys.\ Lett.\ B {\bf 635}, 39 (2006).

\bibitem{atsushi} A.\ Higuchi, Classical Quantum
Gravity {\bf 18}, L139 (2001); {\bf 19}, 599 (2002).

\bibitem{mscalar} E.\ Jung, S.\ H.\ Kim, and D.\ K.\ Park,
Phys.\ Lett.\ B {\bf 602}, 105 (2004).

\bibitem{jungetalnpb} E.\ Jung and D.\ K.\ Park,
Nucl.\ Phys.\ {\bf B717}, 272 (2005).

\bibitem{CHM} L.\ C.\ B.\ Crispino, A.\ Higuchi, and
G.\ E.\ A.\ Matsas, Phys.\ Rev.\ D {\bf 63}, 124008 (2001);
{\bf 80}, 029906(E) (2009).

\bibitem{Gubser} S.\ S.\ Gubser, Phys.\ Rev.\ D {\bf 56}, 7854 (1997).

\bibitem{Gerlach} U.\ H.\ Gerlach,
Phys.\ Rev.\ Lett.\ {\bf 32}, 1023 (1974).

\bibitem{Zerilli} F.\ J.\ Zerilli,
Phys.\ Rev.\ D {\bf 9}, 860 (1974).

\bibitem{Moncrief} V.\ Moncrief,
Phys.\ Rev.\ D {\bf 9}, 2707 (1974);
{\bf 10}, 1057 (1974); {\bf 12}, 1526 (1975).

\bibitem{Olson-Unruh} D.\ W.\ Olson and W.\ G.\ Unruh,
Phys.\ Rev.\ Lett.\ {\bf 33}, 1116 (1974).

\bibitem{Matzner} R.\ A.\ Matzner,
Phys.\ Rev.\ D {\bf 14}, 3274 (1976).

\bibitem{CHO} L.\ C.\ B.\ Crispino, A.\ Higuchi, and
E.\ S.\ Oliveira, Phys.\ Rev.\ D {\bf 80}, 104026 (2009).

\bibitem{CO} L.\ C.\ B.\ Crispino and E.\ S.\ Oliveira,
Phys.\ Rev.\ D {\bf 78}, 024011 (2008).

\bibitem{Ishibashi-Kodama}
H.\ Kodama and A.\ Ishibashi, Prog.\ Theoret.\ Phys.\ {\bf 111}, 29 (2004).

\bibitem{Chandrasekhar} S. Chandrasekhar,
{\it The Mathematical Theory of Black Holes}
(Oxford University Press, New York, 1983).

\bibitem{Myers-Perry} R.\ C.\ Myers and M.\ J.\ Perry, Ann.\ Phys.\ (N.Y.) {\bf 172}, 304 (1986).

\bibitem{CM} A.\ Chodos and E.\ Myers, Ann.\ Phys.\ (N.Y.) {\bf 156}, 412 (1984).

\bibitem{CH} R.\ Camporesi and A.\ Higuchi, J.\ Geom.\ Phys.\ {\bf 15}, 57 (1994).

\end{thebibliography}
\end{document}